\documentclass[usenatbib,iop,numberedappendix]{aeb_emulateapj_2010}
\usepackage{amsmath}
\usepackage{amssymb}

\usepackage{natbib}

\usepackage[usenames,dvips]{color}

\def\s{{\rm s}} 
\def\Ms{{\rm M}\s} 
\def\yr{{\rm yr}} 
\def\hr{{\rm hr}}

\def\GHz{{\rm GHz}} 

\def\m{{\rm m}} 
\def\mm{{\rm m}\m} 
\def\cm{{\rm c}\m} 
\def\km{{\rm k}\m} 
\def\pc{{\rm pc}} 
\def\kpc{{\rm k}\pc} 

\def\Ms{M_\odot} 

\def\erg{{\rm erg}} 

\def\Jy{{\rm Jy}} 
\def\mJy{{\rm m}\Jy} 

\def\AU{{\rm AU}} 

\def\mas{{\rm mas}} 
\def\muas{\mu{\rm as}} 

\renewcommand{\d}{{\rm d}}
\newcommand{\e}{{\rm e}}
\def\Ms{{M_\odot}}
\def\Rs{{r_{\rm S}}}
\def\CARMA{{\rm CARMA}}
\def\SMA{{\rm SMA}}
\def\JCMT{{\rm JCMT}}
\def\SMT{{\rm SMT}}

\def\ALMA{{\rm ALMA}}

\def\BIC{{\rm BIC}}
\def\AIC{{\rm AIC}}
\def\IC{{\rm IC}}

\newcommand\bmath[1] {\mbox{\boldmath$\rm #1$}}

\newcommand\orphanfootnotemark {\addtocounter{footnote}{-1}\footnotemark}

\begin{document}

\title{Evidence for Low Black Hole Spin and Physically Motivated
  Accretion Models from Millimeter VLBI Observations of Sagittarius A*} 

\author{
Avery E.~Broderick\altaffilmark{1}
Vincent L.~Fish\altaffilmark{2}
Sheperd S.~Doeleman\altaffilmark{2}
Abraham Loeb\altaffilmark{3}
}
\altaffiltext{1}{Canadian Institute for Theoretical Astrophysics, 60 St.~George St., Toronto, ON M5S 3H8, Canada; aeb@cita.utoronto.ca}
\altaffiltext{2}{Massachusetts Institute of Technology, Haystack Observatory, Route 40, Westford, MA 01886.}
\altaffiltext{3}{Institute for Theory and Computation, Harvard University, Center for Astrophysics, 60 Garden St., Cambridge, MA 02138.}

\shorttitle{VLBI Parameter Estimation of Sgr A*}
\shortauthors{Broderick et al.}

\begin{abstract}
Millimeter very-long baseline interferometry (mm-VLBI) provides the
novel capacity to probe the emission region of a handful of
supermassive black holes on sub-horizon scales.  For Sagittarius A*
(Sgr A*), the supermassive black hole at the center of the Milky Way,
this provides access to the region in the immediate vicinity of the
horizon.  Broderick et al. (2009) have already shown that by
leveraging spectral and polarization information as well as accretion
theory, it is possible to extract accretion-model parameters
(including black hole spin) from mm-VLBI experiments containing 
only a handful of telescopes.  Here we repeat this analysis with the
most recent mm-VLBI data, considering a class of aligned, radiatively
inefficient accretion flow (RIAF) models.
We find that 
the combined data set rules out symmetric models for Sgr A*'s flux
distribution at the $3.9\sigma$ level, strongly favoring
length-to-width ratios of roughly 2.4:1.  More importantly, we find
that physically motivated accretion flow
models provide a significantly better fit to the mm-VLBI observations than
phenomenological models, at the $2.9\sigma$ level.  This implies
that not only is mm-VLBI presently capable of distinguishing
between potential physical models for Sgr A*'s emission, but further
that it is sensitive to the strong gravitational lensing associated
with the propagation of photons near the black hole.  Based upon this
analysis we find that the most probable magnitude, viewing angle, and
position angle for the black hole spin are 
$a=0.0^{+0.64+0.86}$,
$\theta={68^\circ}^{+5^\circ+9^\circ}_{-20^\circ-28^\circ}$, and
$\xi={-52^\circ}^{+17^\circ+33^\circ}_{-15^\circ-24^\circ}$ east of
north, where the errors quoted are the $1\sigma$ and $2\sigma$
uncertainties.
\end{abstract}


\maketitle

\section{Introduction} \label{I}
Despite being invoked to power a variety of energetic astrophysical
phenomena, the detailed structure and dynamics of black hole accretion
flows remain a central problem in astrophysics.  Moreover, using
electromagnetic observations to probe the structure and dynamics of
the black hole spacetimes requires a substantial understanding of the
physical processes that determine the fate of the accreting matter.
Only recently has it become possible to probe this physics via
large-scale computational simulations.  Nevertheless, ab initio
calculations are beyond our present capability, requiring numerous
simplifying, and in some cases unphysical, assumptions.  This is
evidenced by the number of models proffered to explain the various
properties of accreting black hole candidates.  In turn, this
ambiguity complicates efforts to use electromagnetic observations to
probe the structure and dynamics of the spacetime surrounding the
black hole.

By virtue of its proximity, the supermassive black hole at the center
of the Milky Way, associated with the bright radio point source
Sagittarius A* (Sgr A*), provides an unparalleled opportunity to study
black hole accretion in detail.  For this reason, Sgr A* may serve as
an exemplar of the larger class of supermassive black holes
specifically, and of black holes in general. 
Presently, the best estimates of the mass and distance of Sgr A* come
from the observations of orbiting stars.  These have yielded
$M=4.3\pm0.5\times10^6\,\Ms$ and $D=8.3\pm0.4\,\kpc$, respectively,
where both include the {\em systematic} uncertainties
\citep{Ghez_etal:08,Gill_etal:09a,Gill_etal:09b}.  The mass is
necessarily confined to within the periapse of nearby stars, giving a
maximum radius of roughly $10^2\,\AU\simeq3\times10^3GM/c^2$, ruling
out many extended objects.  These represent the best mass measurement
for any known black hole to date.

In addition to the dynamical observations, a wealth of spectral and
polarization data exists for Sgr A*.  From these it is apparent that
Sgr A* is unlike many active galactic nuclei, being vastly
underluminous, emitting a bolometric luminosity of roughly
$10^{36}\,\erg$, approximately $10^{-9}$ of Eddington.  This is
especially small in light of the considerable amount of gas within the
black hole's sphere of influence, presumably available for accretion
\citep{Loeb-Waxm:07,Cuad-Naya-Mart:07}.  As a result it is widely
accepted that Sgr A*'s accretion flow is qualitatively different from
those in its active analogs, though perhaps indicative of the roughly
90\% of black holes that are presently not in an active phase.

Nevertheless, the existing spectral and polarization data has produced
a canonical set of components all models for Sgr A* include:
populations of thermal and nonthermal electrons, nearly equipartition
magnetic fields.  Less certain is the structure of the emission
region.  This is evidenced by the variety of models that have been
proposed
\citep[e.g.,][]{Nara_etal:98,Blan-Bege:99,Falc-Mark:00,Yuan-Mark-Falc:02,Yuan-Quat-Nara:03,Loeb-Waxm:07}.
Despite being able to reproduce the observed features of Sgr A*, these
differ dramatically in the morphology of the emitting region.  As a
consequence, many of the theoretical ambiguities can be immediately
addressed by direct probes of the spatial distribution of the emitting
plasma surrounding the central supermassive black hole.

The spectrum of Sgr A* peaks near millimeter wavelengths, implying a
transition from optically thick to optically thin emission.  The
location of this emission is currently debated, however the presence
of short-timescale variability at millimeter, near-infrared and X-ray
wavelengths implies that optically thin emission is dominated by
contributions arising in the immediate vicinity of the black hole.
Furthermore, at millimeter wavelengths the blurring due to interstellar
electron scattering is subdominant.  Thus, at wavelengths of
$1.3\,\mm$ and below it is possible to image the emitting region
surrounding Sgr A*.

Even with the strong gravitational lensing in the vicinity of the
horizon, imaging the immediate vicinity of the black hole requires 
extraordinary resolutions.  The silhouette cast by the horizon on the
surrounding emission is roughly $53\pm2\,\muas$\footnote{The mass and
  distance measurements are strongly correlated, with mass scaling
  roughly as $M\propto D^{1.8}$ \citep{Ghez_etal:08}.}.  At the
present time, this resolution is accessible only via
millimeter-wavelength very-long baseline interferometry (mm-VLBI).
VLBI observations of Sgr A* at $1.4\,\mm$ using the Institut de
Radioastronomie Millim\'etrique (IRAM) $30\,\m$ telescope at Pico Veleta and
one of the $15\,\m$ dishes at Plateau de Bure, produced the size estimate of
$110\pm60\,\muas$, with the large uncertainties due to limited
calibration accuracy \citep{Kric_etal:98}. 

The first successful mm-VLBI observation of Sgr A* with Earth-scale
baselines was performed in
April, 2007, during which visibilities were measured on the
$4.6\times10^3\,\km$ baseline between Mauna Kea, Hawaii to Mount
Graham, Arizona \citep{Doel_etal:08}.  By fitting these with a
gaussian model, \citet{Doel_etal:08} found a typical intrinsic source
size of $37^{+5}_{-3}\,\muas$\footnote{We quote the $1\sigma$ errors
  implied by the $3\sigma$ errors reported in \citet{Doel_etal:08}.}
(after correcting for the sub-dominant broadening due to interstellar
electron scattering), smaller than the black hole silhouette.

Since that time a number of groups have analyzed the 2007 mm-VLBI data
using various physically motivated accretion models for the emission
region
\citep{Brod_etal:09,Huan-Taka-Shen:09,Mosc_etal:09,Dext-Agol-Frag-McKi:10},
inferring from these efforts the black hole spin vector.  Despite
finding generally similar results, these have been limited by the lack
of multiple long baseline observations and the limited north-south
coverage obtained.  Recently, a second, and considerably larger set of
mm-VLBI observations have been reported \citep{Fish_etal:10},
providing the opportunity to revisit, and substantially improve,
constraints upon the black hole spin and accretion physics.

Here we report upon the first effort to do this using a physically
motivated accretion model, similar to that described in
\citet{Brod_etal:09}, that fits the known spectral and polarization
properties of Sgr A*.  In addition to improving the resulting parameter
estimation, it is now possible to identify statistical signatures of
both the asymmetry of the image and the importance of the underlying
physics that governs the image morphology.  Section \ref{sec:SoO}
summarizes the full set of mm-VLBI observations we consider.  Section
\ref{sec:VM} describes the models we consider and how the resulting
visibility data is produced.  How models are compared and the
parameter estimates are produced is discussed in Section
\ref{sec:BDA}.  The fitting process and results are presented in
Section \ref{sec:MF}, and our best estimates for the black hole spin
vector can be found in Section \ref{sec:EBHS}.  Section \ref{sec:OFO}
describes the implications for different potential future
observations.  Finally concluding remarks are collected in Section
\ref{sec:C}.

\vspace{1in}
\section{Summary of Millimeter-VLBI Observations} \label{sec:SoO}
\begin{figure*}
  \begin{center}
    \includegraphics[width=\textwidth]{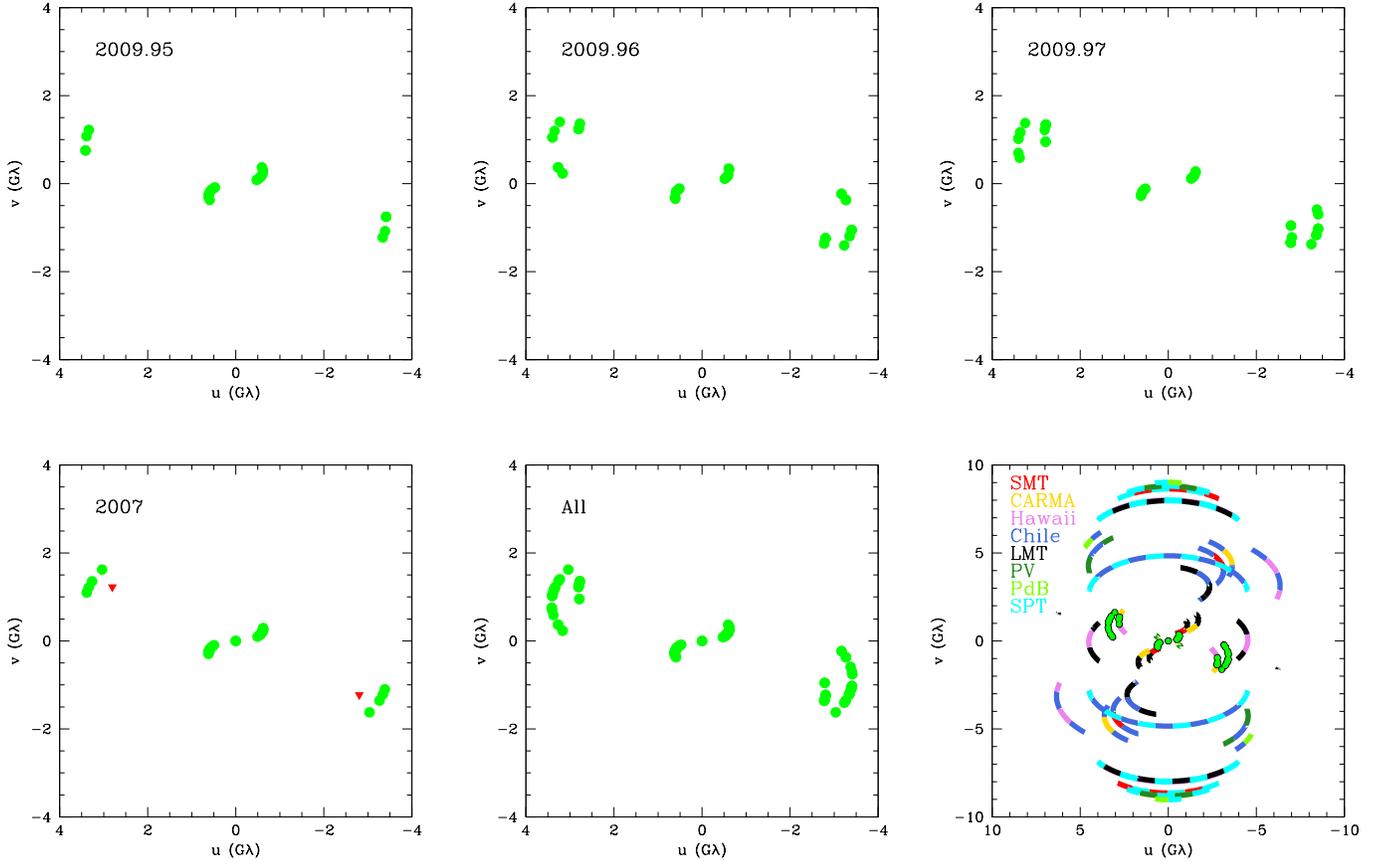}
  \end{center}
  \caption{Locations in the $u$--$v$ plane of the visibilities observed
    during the 2007, 2009.95, 2009.96, 2009.97 epochs.  Also shown are
    the combined set of observations.  Finally, for references the
    combined set is compared to the potential baselines from existing
    and upcoming sub-mm telescopes.  Each baseline is color-coded
    according to the associated two sites.  In all plots, detections are
    denoted by green circles and upper-limits by red triangles.}\label{fig:Vobs}
\end{figure*}
In the analysis presented here we make full use of the recent
observations described in \citet{Fish_etal:10} and
\citet{Doel_etal:08}.  In both cases, observations targeting Sgr A*
were made at $1.3\,\mm$ using the Submillimeter Telescope (\SMT) on
Mt. Graham in Arizona, $10\,\m$ dishes in the Combined Array for
Research in Millimeter-wave Astronomy (\CARMA) at Cedar Flat,
California, and the James Clerk Maxwell Telescope (\JCMT) located on
Mauna Kea, Hawaii.

\subsection{April 2007}
\citet{Doel_etal:08} report upon measurements obtained on the nights
of the April, 11 \& 12, 2007, using the \JCMT, \SMT~and a
single \CARMA~dish.  19 visibility amplitudes were obtained on the
\CARMA--\SMT~and \JCMT--\SMT~baselines, with an upper limit on April
11th, 2007 along the \JCMT--\CARMA~baseline.  The locations of these
observations on the $u$--$v$ plane are indicated in the lower-left
panel of Figure \ref{fig:Vobs}, labeled 2007.  Signal-to-noise ratios
typical of the short and long baselines are 8 and 4, respectively.

During this time, observations the single-dish flux was
estimated via the full \CARMA~array, operating as a stand-alone
instrument, to be $2.4\pm0.25\,\Jy$.  This is similar to the
visibility amplitudes obtained on the \CARMA--\SMT~baselines and
consistent with a single, compact gaussian component
\citep{Doel_etal:08}.  This flux is anomalously low in comparison to
the typical $1.3\,\mm$ flux of $\sim3\,\Jy$, and was taken as evidence
for Sgr A* appearing in a quiescent state.  This interpretation is
supported by the lack of a significant difference between analyses of
each day separately \citep{Brod_etal:09}.

Full details of the observations, calibration and data processing can
be found in \citet{Doel_etal:08}.

\subsection{April 2009}
\citet{Fish_etal:10} report upon more recent observations performed on
the nights of April, 5--7, 2009, corresponding to the 95, 96, and 97
days of 2009.  These made use of the \JCMT, \SMT, and
two \CARMA~dishes, operated as independent VLBI stations.  54
visibility amplitudes were obtained on \JCMT--\SMT~and \CARMA--\SMT~
baselines on all days, and to both of the \JCMT--\CARMA~baselines on
days 96 and 97.  Positions of the observations on each day are
indicated in the upper panels of Figure \ref{fig:Vobs}, labeled
2009.95, 2009.96, and 2009.97.  Signal-to-noise ratios typical of the
short and long baselines are 17 and 5, respectively.  Thus, this
second data set represents a significant improvement in both the
number and precision of the data obtained.

In addition to the VLBI baselines, the presence of two independent
\CARMA~dishes in the array allowed the measurement of very-short
baseline visibilities, probing angular scales $\sim10''$. These found
substantially more correlated flux density than the
\CARMA--\SMT~baselines did, inconsistent with a single compact
gaussian component. The interpretation of the difference in correlated
flux density between the inter-\CARMA~baselines and the
\CARMA--\SMT~baselines is presently unclear, and it may be possible 
for multiple geometric models (e.g., annular rings, extended double
source) to fit the data.  Within the context of our analysis, we will
assume that this difference is due to a separate large-scale component
not present during the 2007 observations.  This is indirectly
supported by the fact that the source sizes inferred from the mid and
long baseline data are unchanged despite the variations in the
visibility magnitudes \citep{Fish_etal:10}.  Therefore, we do not
consider the inter-\CARMA~data further here, restricting ourselves to
modeling the compact component observed with the longer baselines. 

On days 95 and 96 the short-baseline flux densities are consistent
with each other, with inferred single dish fluxes of
$2.15\pm0.06\,\Jy$, which while somewhat lower than those obtained in
2007, justify treating these as a similar quiescent period.  This is
not the case for day 97, which exhibited a $30\%$--$40\%$ increase in
the luminosity of the compact component.  Note that during the
$3\,\hr$ observing periods on days 96 and 97 there is no evidence for rapid
changes in the \CARMA--\SMT~visibility amplitudes, implying that during
each Sgr A* was stable; i.e., the process responsible for the
brightening occurred between observing periods and is stable on
timescales of hours.  As a consequence, we will treat the visibilities
obtained on each day as due to a stationary source, though with
properties that vary from day to day.

Full details of the observations, calibration and data processing can
be found in \citet{Fish_etal:10}.

\subsection{Combined Data Set}
Combined, the 2007 and 2009 mm-VLBI measurements may be separated into
4 observational epochs: that containing the entire 2007 observations (2007),
those on day 95 of 2009 (2009.95), those on day 96 of 2009 (2009.96),
and those on day 97 of 2009 (2009.97).  The combined coverage in the
$u$--$v$ plane is shown in the lower-middle panel of Figure
\ref{fig:Vobs}.  The long baselines (\JCMT--\CARMA~and \JCMT--\SMT)
are oriented primarily in the east-west direction, extended roughly
$3.6\,{\rm G\lambda}$.  Nevertheless, the combined data set also
extends roughly $2\,{\rm G\lambda}$ in the north-south direction,
providing substantial angular coverage in the $u$--$v$ plane for the
first time.

In Section \ref{sec:OFO} we will discuss the implications out analysis
has for future observations.  However, we note here that the baselines
considered in the 2007 and 2009 mm-VLBI experiments are a small subset
of the baselines that are possible with existing mm and sub-mm
telescopes.  Figure \ref{fig:Vobs} shows the combined visibility data
set in comparison to baselines associated with other potential mm-VLBI
stations.  These include stations in Chile (e.g., the Atacama
Pathfinder EXperiment, Atacama Submillimeter Telescope
Experiment, and Atacama Large Millimeter Array; APEX, ASTE, and ALMA,
respectively), Mexico (Large Millimeter Telescope; LMT), Spain (Pico
Veleta; PV), France (Plateau de Bure; PdB), and at the South Pole
(South Pole Telescope; SPT).  These both, extend the region covered in
the $u$--$v$ plane, and provide additional complementary short and
intermediate baselines, primarily along the north-south directions.
To date, visibilities on only a handful of potential baselines have
been measured.

\section{Visibility Modeling} \label{sec:VM}
Our primary goal is to use physically motivated models of Sgr A*'s
accretion flow to infer the properties of the central supermassive
black hole and its surrounding matter.  To do this we compare both
physical and phenomenological models of Sgr A* to the mm-VLBI
visibilities.  This requires the computation of model visibilities.
Given a trial image intensity distribution, $I(\alpha,\beta)$, where
$\alpha$ and $\beta$ are angular coordinates, we may compute the
visibilities in the standard fashion: 
\begin{equation}
V(u,v) = \int\int \d\alpha \d\beta\, e^{-2\pi i(\alpha u + \beta v)/\lambda} I(\alpha,\beta)\,.
\end{equation}
Here we describe three classes of model images: those associated with
radiatively inefficient accretion flows (RIAFs) of the form discussed
in \citet{Brod-Loeb:06a}, symmetric and asymmetric gaussians.  We also
summarize the effects of interstellar electron scattering.

\subsection{Radiatively Inefficient Accretion Flows} \label{sec:RIAF}
We employ a suite of radiatively inefficient accretion flow (RIAF)
models, first described in \citet{Brod-Loeb:06a}, and based upon those
of \citet{Yuan-Quat-Nara:03}.  Here these models, which henceforth we
refer to as BL06, are summarized.

Sgr A* transitions from an inverted, presumably optically thick
spectrum to an optically thin spectrum near millimeter wavelengths.
This implies that near 1.3mm Sgr A* is only becoming optically thin,
and thus absorption in the surrounding medium is likely to be
important.  This transition does not occur isotropically, happening at
longer wavelengths for gas that is receding and at shorter wavelengths
for gas that is approaching.  Therefore, properly modeling the
structure and relativistic radiative transfer is crucial to producing
high fidelity images.

Although Sgr A* is vastly sub-Eddington, its bolometric luminosity,
roughly $10^{36}\,\erg\s^{-1}$, is still large in absolute terms, 
Like many AGN, in the radio Sgr A* exhibits the nearly-flat, power-law
spectrum associated with non-thermal synchrotron sources, with the
power emitted ($\nu L_\nu$) peaking at millimeter wavelengths.  As a
consequence, it has been widely accepted that Sgr A* is accretion
powered, implying a minimum accretion rate of $10^{-10}\Ms\yr^{-1}$.
It is presently unclear how this emission is produced, evidenced by
the variety of models that have been proposed
\citep[e.g.,][]{Nara_etal:98,Blan-Bege:99,Falc-Mark:00,Yuan-Mark-Falc:02,Yuan-Quat-Nara:03,Loeb-Waxm:07}.
Models in which the emission arises directly from the accreting gas
have been subsumed into the general class of RIAFs, defined by the
generally weak coupling between the electrons, which radiate rapidly,
and the ions, which efficiently convert gravitational potential energy
into heat \citep{Nara_etal:98}.  This coupling may be sufficiently
weak to allow accretion rates substantially in excess of that required
to explain the observed luminosity with a canonical AGN radiative
efficiency of $10\%$.  However, the detection of linear polarization
in Sgr A* above $100\GHz$ 
\citep{Aitk_etal:00,Bowe_etal:01,Bowe_etal:03,Marr_etal:06}
and subsequent measurements of the Faraday rotation measure
\citep{Macq_etal:06,Marr_etal:07},
have implied that the accretion rate near the black hole is much less
than the Bondi rate, requiring the existence of large-scale outflows
\citep{Agol:00,Quat-Gruz:00}.

\begin{figure}
  \begin{center}
    \includegraphics[width=\columnwidth]{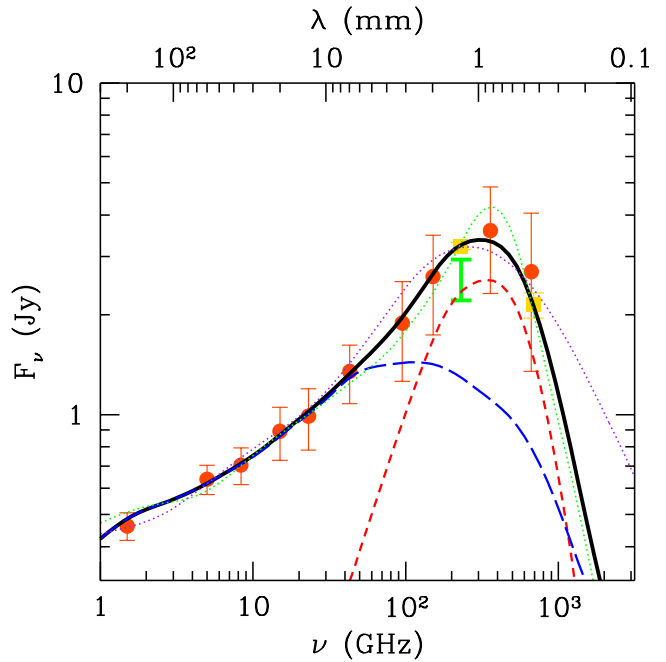}
  \end{center}
  \caption{Comparison of the spectrum of the most probable accretion model
    and the observed SED of Sgr A*.  Orange circles are from
    \citet{Yuan-Quat-Nara:04}, and references therein, for
    which the errorbars are indicative of the variability.  Yellow
    squares are coincident flux measurements from \citet{Marr:06}, for
    which the errorbars are indicative of the intrinsic measurement
    error.  The green bar shows the flux range of the compact
    component inferred from fitting accretion models to the mm-VLBI
    data (Section \ref{sec:MF}.  In addition to the most probable
    model, spectra are shown for emission from a rapidly rotating
    black hole ($a=0.998$) as seen nearly face on ($\theta=1^\circ$,
    green dotted) and edge on ($\theta=90^\circ$, purple dotted),
    indicating the range of variation within the image library.
    Finally, for reference the
    contribution from the thermal (red dash) and nonthermal (blue
    long-dash) are shown.  Note in particular that at 1.3mm the
    nonthermal contribution is not negligible. }\label{fig:spec}
\end{figure}

\begin{figure*}
  \begin{center}
    \includegraphics[width=\textwidth]{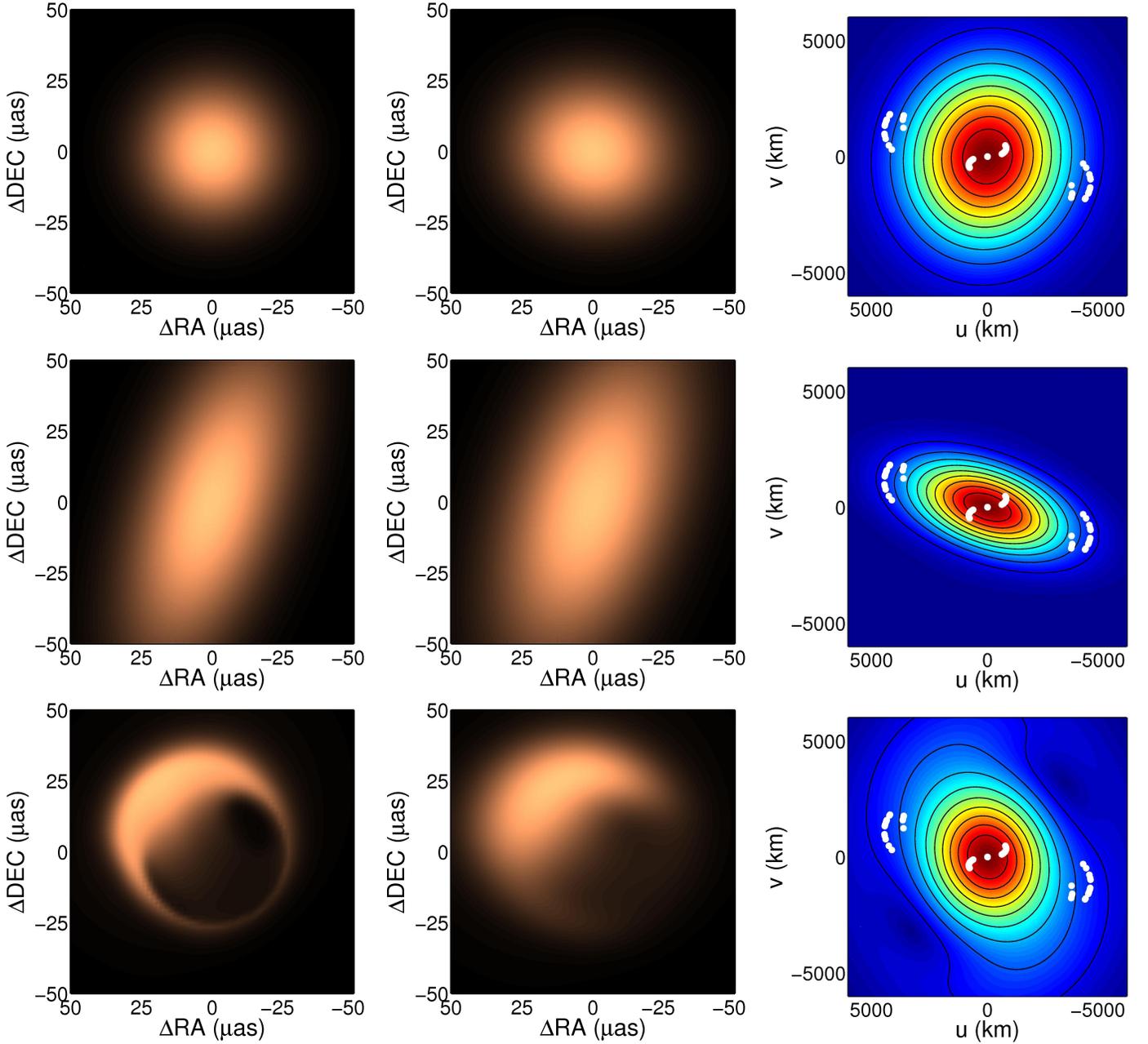}
  \end{center}
  \caption{Best fit images and visibilities for the three models
    considered: symmetric gaussian (top), asymmetric gaussian (middle)
    and BL06 accretion flow (bottom).  For each model we show the
    intrinsic flux distribution (left), flux distribution after
    interstellar electron scattering (center) and the visibility
    amplitudes (right).  For reference the locations of the observed
    visibilities (over all epochs) are shown by the white points.  In
    all plots the intensity scales linearly with $I$ and $V$.}
  \label{fig:bestfit}
\end{figure*}

Relating the outflow to the properties of the accretion flow requires
an ab initio calculation that is presently not possible.
Nevertheless, a number of authors have studied this relationship in
the context of a variety of simplifying assumptions, with large-scale
general-relativistic magnetohydrodynamic and radiative-hydrodynamic simulations
playing a central role
\citep{DeVi-Hawl-Krol-Hiro:05,McKi:06,Hawl-Krol:06,Beck-Hawl-Krol:08,McKi-Blan:09,Tche-Nara-McKi:10,Dext-Agol-Frag-McKi:10,Penn_etal:10,Kuro-Prog:09}.
In these it has been found that the
structure and dynamics of the outflow critically depends upon the
initial conditions.  The applicability of the MHD prescription
to Sgr A* is still unclear, where the accretion rate is sufficiently
low that non-MHD effects may become important
\citep{Shar-Hamm-Quat-Ston:06,Shar-Quat-Hamm-Ston:07}.  More
importantly, most of these approaches do not model the electron
heating (beyond ad hoc prescriptions) and none model the production of
nonthermal electrons
\citep[see, e.g.,][]{Mosc_etal:09,Dext-Agol-Frag-McKi:10,Shch-Penn-McKi:10}.  
Furthermore, simulations are computationally expensive to produce.
For these reasons we adopt a simple, self-similar model for the
accretion flow which includes substantial mass loss.

For concreteness, as in \citet{Brod-Loeb:06a}, we follow
\citet{Yuan-Quat-Nara:03} and employ a model in which the accretion
flow has a Keplerian velocity distribution, a population of thermal
electrons with density and temperature
\begin{equation}
n_{e,{\rm th}}=n_{e,{\rm th}}^0 \left(\frac{r}{\Rs}\right)^{-1.1} e^{-z^2/2\rho^2}
\end{equation}
and
\begin{equation}
T_{e}=n_{e}^0 \left(\frac{r}{\Rs}\right)^{-0.84}\,,
\end{equation}
respectively,
and a toroidal magnetic field in approximate ($\beta=10$)
equipartition with the ions (which are responsible for the majority of
the pressure), i.e.,
\begin{equation}
\frac{B^2}{8\pi} = \beta^{-1} n_{e,{\rm th}} \frac{m_p c^2 \Rs}{12 r}\,.
\end{equation}
In all of these, $\Rs=2GM/c^2$ is the Schwarzschild radius, $\rho$ is
the cylindrical radius and $z$ is the vertical coordinate.  Inside of
the innermost-stable circular orbit (ISCO) we assume the gas is
plunging upon ballistic trajectories.  In principle the plunging gas
can still radiate, though in practice it contributes little to the
overall emission due to the large radial velocities it develops.  In
the case of the thermal quantities the radial structure was taken from
\citet{Yuan-Quat-Nara:03}, and the vertical structure was determined
by assuming that the disk height is comparable to $\rho$.  Note that
all of the models we employ necessarily have the spin aligned with the
orbital angular momentum of the accretion flow.  For the regions that
dominate the $\mm$ emission, this assumption is well justified due to
disk precession and viscous torques, though it may be violated at
large distances.

Thermal electrons alone are incapable of reproducing the nearly-flat
spectrum of Sgr A* below $43\,\GHz$.  Thus it is necessary to also
include a nonthermal component.  As with the thermal components, we
adopt a self-similar model for a population of nonthermal electrons,
\begin{equation}
n_{e,{\rm nth}}=n_{e,{\rm nth}}^0 \left(\frac{r}{\Rs}\right)^{-2.02} e^{-z^2/2\rho^2}\,,
\end{equation}
with a power-law distribution corresponding to a spectral index of
$1.25$ and cut off below Lorentz factors of $10^2$ \citep[consistent with
][]{Yuan-Quat-Nara:03}.  The radial power-law index was chosen to
reproduce the low frequency spectrum of Sgr A*, and is insensitive to
the black hole properties due to the distant location of the
long-wavelength emission.

The primary emission mechanism at the wavelengths of interest is
synchrotron, arising from both the thermal and nonthermal electrons.
We model the emission from the thermal electrons using the emissivity
described in \citet{Yuan-Quat-Nara:03}, appropriately altered to
account for relativistic effects \citet[see, e.g., ][]{Brod-Blan:04}.
Since we perform polarized radiative transfer via the entire
complement of Stokes parameters, we employ the polarization fraction
for thermal synchrotron as derived in \citet{Petr-McTi:83}.  In doing
so, we have implicitly assumed that the emission due to thermal
electrons is isotropic, which while generally not the case is unlikely
to change our results significantly.  For the nonthermal electrons, we
follow \citet{Jone-ODel:77} for a power-law electron distribution,
with an additional spectral break associated with the minimum electron
Lorentz factor.  For both emission components the absorption
coefficients are determined directly via Kirchhoff's law.  Images are
then produced using the fully relativistic ray-tracing and radiative
transfer schemes described in \citet{Brod-Loeb:06a,Brod-Loeb:06b} and
\citet{Brod:06}.  An example image and associated visibilities are
shown in the bottom line of Figure \ref{fig:bestfit}.

Because \citet{Yuan-Quat-Nara:03} neglected relativistic effects and
assumed spherical symmetry, it is not directly applicable here.  For
these reasons, as in \citet{Brod-Loeb:06a}, the coefficients
$(n_{e,{\rm th}}^0,T_e^0,n_{e,{\rm nth}}^0)$ were adjusted to fit the
radio spectral energy distribution (SED) of Sgr A*, shown in Figure
\ref{fig:spec}.  The $1.3\,\mm$ and $0.43\,\mm$ points (orange
squares) were measured coincidentally, and the errors represent the
intrinsic measurement error \citep{Marr:06}.  All other points (red
circles) are taken from \citet{Yuan-Quat-Nara:04} (and references
therein) and were obtained by averaging over multiple epochs.  As a
result the errorbars represent the range of variability, and are
correspondingly larger.

As in \citet{Brod_etal:09} we systematically
fit Sgr A*'s SED at a large number of positions in the
spin-inclination ($a$-$\theta$, where $\theta$ is the viewing angle
relative to the disk axis) parameter space, specifically at
values of $a\in\{0,0.1,0.2, ... ,0.9,0.99,0.998\}$ for all
$\theta\in\{1^\circ,10^\circ,20^\circ,...,80^\circ,90^\circ\}$,
producing a tabulated set of the coefficients 
$(n_{e,{\rm th}}^0,T_e^0,n_{e,{\rm nth}}^0)$ at
120 points in the $a$-$\theta$ parameter space.  In all cases it was
possible to fit the SED with extraordinary precision, with reduced
$\chi^2<0.3$ in all cases and typically reduced $\chi^2\simeq0.17$.
This is likely a consequence of employing the variability-determined
errorbars on the non-coincident flux measurements.  Over the
$a$-$\theta$ plane the $\chi^2$ was remarkably uniform, implying that
on the basis of the spectra alone it is difficult to constrain the
parameters of our simple model.  Nevertheless, 
the dynamical properties of the disk manifest themselves in the
breadth of the sub-millimeter bump.  Models with low spin and/or low
$\theta$ have little Doppler shifting, and correspondingly narrow
bumps.  In contrast, models with large spins ($a>0.9$) viewed
edge-on ($\theta>80^\circ$) have broad bumps, and are responsible for
the relatively larger, though still small, $\chi^2$.  From the
tabulated values, the coefficients are then obtained at arbitrary $a$
and $\theta$ using high-order polynomial interpolation.

An example spectrum, resulting from the above procedure, is shown in
Figure \ref{fig:spec}, corresponding to $a=0.01$ and
$\theta=66^\circ$.  In addition, the individual contributions from the
thermal and nonthermal components are also shown (though the
cross-absorption is neglected).  The necessity of the nonthermal
electrons is clearly illustrated at long wavelengths, where the
thermal contribution is negligible.  The transition from nonthermally
dominated to thermally dominated occurs near $2\,\mm$, though the
precise location depends upon $a$ and $\theta$.  However, note that at
no point can the nonthermal component be neglected.  Specifically, at
$1.3\,\mm$ the nonthermal component is still responsible for roughly
$30\%$ of the emission.  As a result, efforts to model the millimeter
image of Sgr A* without accounting for the nonthermal component can
produce order unity systematic errors in parameter estimation.

During the mm-VLBI observations Sgr A*'s flux varied by roughly
$30\%$.  We model this as a variable accretion rate, moving the
electron density normalization up and down.  In practice, we reduced
the electron density normalization by an amount sufficient to produce
a total flux of $2.5\,\Jy$, and then multiplied the resulting images
by a correction faction during the mm-VLBI data analysis.  Because the
source is not uniformly optically thin, this is not strictly correct,
though this makes a small change to the images themselves.  For the
purpose of the mm-VLBI data analysis (described below) we produced
9090 images, with flux normalized as described above, at
$a\in\{0,0.01,0.02,...,0.98,0.99,0.998\}$ for each
$\theta\in\{1^\circ,2^\circ,...,89^\circ,90^\circ\}$.  We then produce
models with arbitrary position angles, $\xi$, by rotating the image on
the sky.  For this purpose we define $\xi$ such that at $0^\circ$ the
projected spin vector points north, and as $\xi$ increases points
progressively more eastward\footnote{Note that this is opposite the
  definition employed in \citet{Brod_etal:09}.}.

The Faraday rotation measures observed in Sgr A* are produced at much
larger radii than those of interest in direct imaging experiments.
Nevertheless, it is worth noting that the models employed here are 
broadly consistent with the polarization observations, though breaks
in the radial power-laws which define the properties of the thermal
electron component may be required at large spins.

For the purposes of fitting the mm-VLBI visibilities, during each
observation epoch this model has 4 parameters: spin ($a$), viewing
angle ($\theta$), position angle ($\xi$) and flux normalization
($V_{00}$).  When we analyze multiple epochs together the parameters
defining the orientation of the system ($a$,$\theta$,$\xi$) will be
held fixed, while those corresponding to the time-variable accretion
rate ($V_{00}$) will be allowed to vary, though a full discussion will
have to await Section \ref{sec:MF}.

\subsection{Gaussian Flux Distributions}
For comparison we consider two gaussian flux distributions.  These
differ from the accretion flow model described above in that they are
purely phenomenological, without any clear physical motivation and
thus not constrained at all by the spectral and polarization
properties of Sgr A*.  As a result, we might expect these to be
intrinsically less likely than physically motivated models that are
already chosen to be consistent with these properties.  Nevertheless,
we will consider them on equal footing with the BL06 model discussed
above.  For reasons that will become clear, we consider both symmetric
and asymmetric gaussian intrinsic flux distributions.

We may write the asymmetric gaussian flux distribution as
\begin{equation}
I = V_{00} \exp\left(-\frac{\alpha_M^2}{2\sigma_M^2}-\frac{\alpha_m^2}{2\sigma_m^2}\right)\,,
\end{equation}
where $\alpha_{M,m}$ and $\sigma_{M,m}$ are the angular coordinates
and widths in the major/minor axis directions.
This is
fully defined once $\sigma_{M,m}$ and the position angle of the major
axis is given.  However, we choose to parametrize the asymmetric
gaussian in terms of a single width, $\sigma$, an anisotropy
parameter, $A$, and the position angle:
\begin{equation}
I = V_{00} \exp\left(-\frac{\alpha^2}{2\sigma^2} - A\frac{\alpha^2}{2\sigma^2}\cos2\varpi\right)\,,
\end{equation}
where $\alpha=\sqrt{\alpha_M^2+\alpha_m^2}$ and $\varpi$ is the
angular coordinate measured from the position angle.  The
$\sigma$ and $A$ are related to the $\sigma_{M,m}$ by
\begin{equation}
\frac{1}{\sigma^2} = \frac{1}{2\sigma_m^2}+\frac{1}{2\sigma_M^2}
\quad{\rm and}\quad
\frac{A}{\sigma^2} = \frac{1}{2\sigma_m^2}-\frac{1}{2\sigma_M^2}\,.
\end{equation}
Example gaussian images, with associated visibilities, are shown in
the top two lines of Figure \ref{fig:bestfit}.

Clearly, for isotropic configurations (i.e., $\sigma_M=\sigma_m$),
$A=0$ and $\sigma=\sigma_{M,m}$.  More generally,
$\sigma_m/\sigma_M = \sqrt{(1-A)/(1+A)}$.  Thus, this parametrization
has the virtue of separately emphasizing size (via $\sigma$) and
asymmetry (via $A$) in the image.  Note that this model has precisely
the number of free parameters as the accretion flow model described
above: those describing the image morphology,
$(\sigma,A,\xi)$, and the flux normalization, $V_{00}$, for each epoch.

While the symmetric gaussian models are obviously a subset of the
asymmetric gaussian models (corresponding to when $A$ vanishes), we
must be careful to distinguish the number of free parameters.  In this
case, $A$ and $\xi$ are superfluous, and for each epoch we have only 1
parameter.

\subsection{Interstellar Electron Scattering}
The effect of interstellar electron scattering in the direction of
Sgr A* have been carefully characterized empirically by a number of
authors.  This has been found to be consistent with convolving the
source with an asymmetric gaussian, with major axis nearly aligned
with east-west, and a $\lambda^2$ wavelength dependence.  We employ
the model from \citet{Bowe_etal:06}, which has major axis oriented
$78^\circ$ east of north, with associated full width at half-maximum
for the major and minor axes given by
\begin{equation}
\begin{aligned}
{\rm FWHM}^{\rm ES}_M &= 1.309\left(\frac{\lambda}{1\,\cm}\right)^{2}\,\mas\,,\\
{\rm FWHM}^{\rm ES}_m &= 0.64\left(\frac{\lambda}{1\,\cm}\right)^{2}\,\mas\,,
\end{aligned}
\end{equation}
respectively.  In practice, the interstellar electron scattering
convolution was effected in the $u$--$v$ plane, where the convolution
reduces to a multiplication.

\section{Bayesian Data Analysis} \label{sec:BDA}
In fitting the observed visibilities  we have two primary goals:
choosing among various possible model flux distributions and
estimating the parameters of these models.  Both of these may be
naturally accomplished within the context of Bayesian analysis.  Here
we briefly summarize how we do this.

We define the likelihood, $p(\bmath{V}|\bmath{q})$, for observing the
visibilities $\bmath{V}$ given the model parameters $\bmath{q}$, as
described in \citet{Brod_etal:09}.  From this we obtain the
log-likelihood, which we refer to as $\chi^2$:
\begin{equation}
\chi^2 \equiv -2 \log p(\bmath{V}|\bmath{q}) + C\,,
\end{equation}
where the normalization constant depends only upon the particulars of
the data and, since we will only be interested in comparing identical data
sets, will henceforth be ignored.  When only detections are considered
this reduces trivially to the standard definition of $\chi^2$.  Here
it differs only due to the presence of an upper-limit upon the
visibility along the \CARMA-\JCMT~baseline during the 2007 epoch.

\subsection{Model Comparison}
To compare the significance of different models we make use of the
Bayesian Information Criterion ($\BIC$) and the Akaike Information
Criterion ($\AIC$).  Both of these are discussed in detail within the
context of astrophysical observations by \citet{Lidd:07} and
\citet{Take:00}, and references therein.  Thus, here we only define
and summarize the properties of these statistics.

In terms of the smallest effective $\chi^2$ for a given model, the
$\BIC$ is defined by
\begin{equation}
\BIC \equiv \chi_{\rm min}^2 + k \ln N\,,
\end{equation}
where $k$ is the number of model parameters and $N$ is the number of
data points.  Note that this is simply the $\chi^2$ statistic
penalized for models with large numbers of parameters.  Assuming that
the data points are independent (likely true) and identically
distributed (nearly true), this is related to the posterior
probabilities of two different models, $M_1$ and $M_2$, by 
\begin{equation}
\frac{p(M_1|\bmath{V})}{p(M_2|\bmath{V})}
=
\frac{p(M_1)}{p(M_2)} \e^{-(\BIC_1-\BIC_2)/2}\,,
\end{equation}
where $p(M_{1,2})$ is the prior on model $M_{1,2}$.  Thus, up to the
unknown priors, the $\BIC$ is a measure of the posterior probability
for a given model.  If we further assume that $p(M_1)=p(M_2)$, the
difference in $\BIC$'s gives the relative posterior probabilities
directly.  Therefore the model with the lowest $\BIC$ is preferred.

Similarly, the $\AIC$ is defined by
\begin{equation}
\AIC \equiv \chi_{\rm min}^2 + 2 k + \frac{2 k (k+1)}{N-k-1}
\end{equation}
where we have included a correction appropriate for when $N$ is small
\citep{Burn-Ande:02,Burn-Ande:04}.  This is very similar to the
$\BIC$: the $\chi^2$ statistic penalized by a factor depending upon
the number of model parameters, though with a somewhat different
penalty.  Unlike the $\BIC$, the $\AIC$ is not directly related to the
posterior probability of a given model.  Rather it is an approximate
measure of the difference between the true data distribution and the
modeled data distribution.  Nevertheless, it is possible to interpret
the $\AIC$ in terms of a model likelihood in a way identical to the
$\BIC$.

For either criterion, lower values are preferred, with the relative
significance given by
\begin{equation}
w_{ij} = \e^{-(\IC_i-\IC_j)/2}\,,
\end{equation}
where $\IC$ may be replaced by $\BIC$ or $\AIC$.  The $\Delta\IC$ are conventionally
judged on the Jeffrey's scale, which sets $\Delta\IC>5$ as ``strong''
and $\Delta\IC>10$ as ``decisive'' evidence against the model with the
higher $\IC$.  However, here we describe these in terms of the typical
$\sigma$ as well, with model $i$ being excluded at the $n\sigma$
level if there exists a model $j$ for which $w_{ij}$ is less than the 
associated cumulative normal probability (e.g., $1\sigma$ implies
that $w_{ij}<0.32$, $2\sigma$ implies that $w_{ij}<0.05$,
$3\sigma$ implies that $w_{ij}<0.003$, etc.).

\subsection{Parameter Estimation}
For the accretion flow model we have the additional problem of
identifying the most likely model parameters.  The procedure we use to
estimate the posterior probabilities is identical to that described in
\citet{Brod_etal:09}.  In particular, we assume flat priors on all of
the visibility normalizations, $\bmath{V_{00}}$, the spin magnitude,
$a$, and an isotropic prior on the spin direction $(\theta,\xi)$.
Since we are primarily interested in the estimates for the black hole
spin, we present the posterior probability of $\bmath{a}$ marginalized
over the $\bmath{V_{00}}$, $p(\bmath{a})$.  We also
construct marginalized posterior probability distributions of $a$,
$\theta$, and $\xi$ in the normal way
\citep[for specific definitions of $p(a)$, $p(\theta)$, and $p(\xi)$, see][]{Brod_etal:09}.

\section{Model Fitting} \label{sec:MF}
A number of important implications follow from computations of the
relevant $\chi^2$ for the three image models described in Section
\ref{sec:VM}.  These include whether or not we are justified in
comparing mm-VLBI observations obtained at different times, the
symmetry of the image and the importance of physics for reproducing
the measured visibilities.

\begin{deluxetable*}{ccccccccccccc}\tabletypesize{\small}
\tablecaption{Model Fitting Results Summary\label{tab:results}}
\tablehead{
\colhead{Model} &
\colhead{$k$} &
\colhead{$\chi^2$} &
\colhead{${\rm DoF}$} &
\colhead{$\chi^2/{\rm DoF}$} &
\colhead{$V_{00}^{2007}$} &
\colhead{$V_{00}^{2009.95}$} &
\colhead{$V_{00}^{2009.96}$} &
\colhead{$V_{00}^{2009.97}$} &
\colhead{$\BIC$} &
\colhead{$w^\BIC_{i,BL06}$} &
\colhead{$\AIC$} &
\colhead{$w^\AIC_{i,BL06}$}
}
\startdata
\cutinhead{Estimated Errors}
Symmetric Gaussian  & 5 & 76.78 & 66 & 1.16  & 2.37 & 2.08 & 2.03 & 2.88 & 98.1  & $5\times10^{-4}$ & 87.7 & $8\times10^{-5}$\\
Asymmetric Gaussian & 7 & 61.50 & 64 & 0.961 & 2.53 & 2.25 & 2.23 & 3.06 & 91.3  & $1\times10^{-2}$ & 77.3 & $1\times10^{-2}$\\
BL06                & 7 & 53.09 & 64 & 0.830 & 2.45 & 2.18 & 2.16 & 3.00 & 82.9  & 1                & 68.9 & 1\\
\hline
\cutinhead{Implied Errors}
Symmetric Gaussian  & 5 & 92.51 & 66 & 1.40  & 2.37 & 2.08 & 2.03 & 2.88 & 114   & $4\times10^{-5}$  & 103  & $9\times10^{-6}$\\
Asymmetric Gaussian & 7 & 74.10 & 64 & 1.16  & 2.53 & 2.25 & 2.23 & 3.06 & 104   & $6\times10^{-3}$  & 89.9 & $6\times10^{-3}$\\
BL06                & 7 & 63.97 & 64 & 1.00  & 2.45 & 2.18 & 2.16 & 3.00 & 93.8  & 1                 & 79.7 & 1
\enddata
\end{deluxetable*}

\subsection{Consistency of the 2007 \& 2009 Epochs}
The dynamical timescale of the Sgr A* is comparable to the orbital
period at the ISCO, as measured at Earth, and ranges from roughly
$4\,\min$ to $30\,\min$, depending upon spin.  As a result it is not
at all clear that we may ignore the possibility of variability when
attempting to model mm-VLBI observations spanning many nights, let
alone years.  \citet{Brod_etal:09} took special pains to ensure that
the visibilities measured on the two consecutive days were consistent
with a single, static underlying flux distribution.  This was done by
comparing fits to the individual days.  Here we repeat this analysis
for the 2009 observations as well.

In 2007, at $2.4\,\Jy$, the luminosity of Sgr A* was anomalously low
and stable over the two observation days.  This is not the case during
the 2009 observation, during which Sgr A* exhibited a dramatic
brightening on the third day.  During the preceding two days the
luminosity of Sgr A*'s compact component, corresponding to scales
smaller than $10^2\Rs$, was significantly smaller than that associated
with the 2007 observations.  Thus it is clear from the outset that we
are not justified in comparing a single, static model to the
observations.  Instead, we begin with the ansatz that the morphology
of Sgr A*'s image is fixed, with the flux variability being driven by
changes in the accretion rate on day-to-day timescales.  While this
period is considerably larger than the $30\,\min$ timescale over which
the properties of the accretion flow may change, it is justified in
part by the stability of Sgr A*'s luminosity on these scales as well
as the intrinsically short duration ($\sim2\,\hr$) of the
observations each night.
Therefore, we separate the data into 4 epochs: 2007, 2009.95, 2009.96
and 2009.97, corresponding to the data obtained in 2007 and on days
95, 96 and 97 of 2009, respectively.  For all epochs we keep the
parameters that define the image morphology fixed, e.g.,
$(a,\theta,\xi)$, $(\sigma,A,\xi)$, or $\sigma$, but allow the overall
flux normalization to vary from epoch to epoch.  Upon fitting each
epoch separately, and all epochs together, we may ask if the resulting
parameter likelihoods are consistent with each other, i.e., check if
our ansatz is self consistent.  We will remark upon this further in
the sections describing the fits for the individual models; however
here it is sufficient to note that in all cases we find that the
epochs are indeed consistent with a single underlying image
morphology.

\subsection{Symmetric Gaussian}
\begin{figure}
  \begin{center}
    \includegraphics[width=\columnwidth]{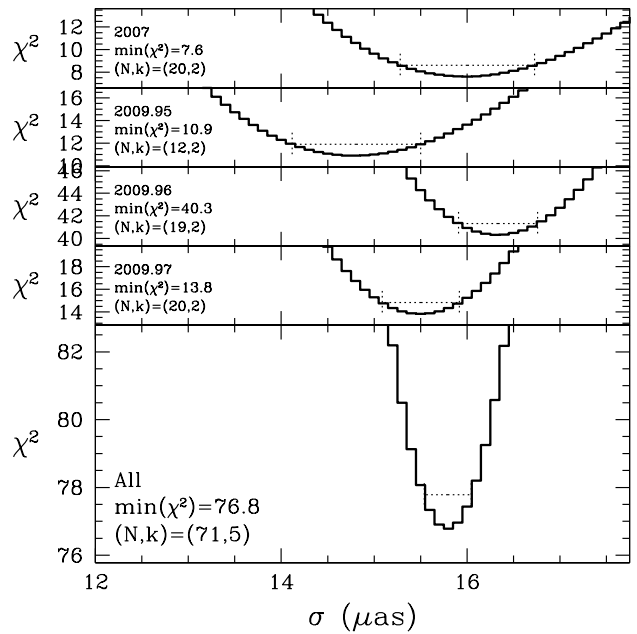}
  \end{center}
  \caption{Symmetric gaussian $\chi^2$'s as a function of $\sigma$, at
    the most likely $\bmath{V_{00}}$, for the
    2007, 2009.95, 2009.96, and 2009.97 epochs, as well as when all
    epochs are combined.  The minimum $\chi^2$, number of visibility
    observations ($N$) and fit parameters ($k$) are listed in each plot.
    In all cases the range shown corresponds to $\min(\chi^2)-1$ to
    $\min(\chi^2)+6$.  For references the $1\sigma$ error estimate is
    shown by the dotted errorbars for each data set.}\label{fig:sgchi2}
\end{figure}
We begin with the symmetric gaussian model, which is primarily
sensitive to the characteristic size of the image.  In this case for
each epoch there are 2 parameters: $\sigma$ and $V_{00}$.  The
minimum $\chi^2$ is shown as a function of $\sigma$ for each epoch in
Figure \ref{fig:sgchi2}.  With the exception of epoch 2009.96, the
reduced-$\chi^2$ is comparable to unity.  Over all epochs the $\sigma$
with the highest likelihood varies over $1.5\,\muas$, well within the
single-epoch uncertainty (defined by the region in which $\chi^2$ is
within unity of the minimum value).  In particular, there are no
trends distinguishing either the much earlier 2007 epoch or the
considerably brighter 2009.97 epoch.  Thus we conclude that the
characteristic size of Sgr A* did not vary substantially from one
epoch to the next, despite considerably changes in its luminosity.

The best fit intrinsic source size is $\sigma = 15.8\pm0.2\,\muas$
(${\rm FWHM}=37.2\pm0.5\,\muas$) with the flux normalization provided in
Table \ref{tab:results}, ranging from $2.07\,\Jy$ to $2.81\,\Jy$ over
the various epochs.   The associated reduced-$\chi^2$ is $1.16$, with
$66$ degrees of freedom.  Upon convolving with the interstellar
electron scattering, which along the general direction of the
baselines \SMT--\JCMT~and \CARMA--\JCMT~baselines has a ${\rm FWHM}$ of
width of $22\,\muas$, we find the ${\rm FWHM}$ of the broadened image
is $43.2\pm0.6\,\muas$.  This is in excellent agreement with the
inferred size of $43^{+5}_{-3}$ found by \citet{Doel_etal:08} on the
basis of the 2007 epoch alone\footnote{Here we have quoted the
  $1\sigma$ errors upon their result in order to provide a direct
  comparison.}.  It is also in excellent agreement with the inferred
sizes of $41.3^{+1.8}_{-1.4}$, $44.4^{+1}_{-1}$, and
$42.6^{+1}_{-1}$ found by \citet{Fish_etal:10} for epochs 2009.95,
2009.96, and 2009.97, respectively\orphanfootnotemark.  The associated intrinsic and
scatter-broadened image, with the associated visibilities are shown in
the left, center and right panels of the top row of Figure
\ref{fig:bestfit}.

\subsection{Asymmetric Gaussian and Image Anisotropy}
\begin{figure*}
  \begin{center}
    \includegraphics[width=\textwidth]{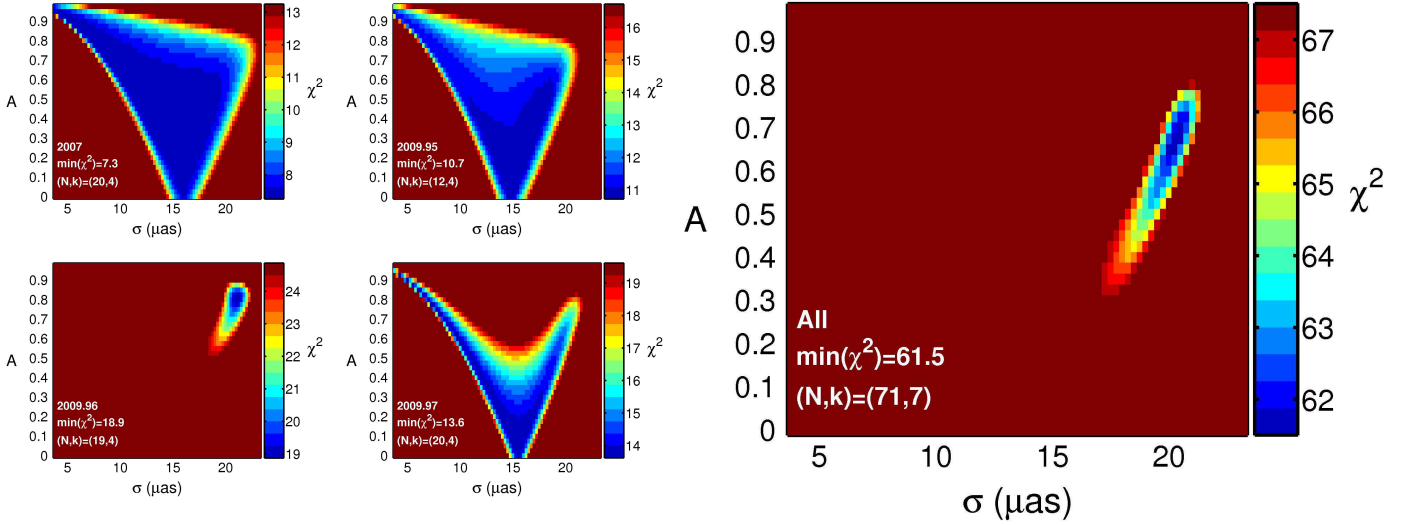}
  \end{center}
  \caption{Asymmetric gaussian $\chi^2$'s as a function of $\sigma$ and
    $A$, at the most likely $\xi$ and $\bmath{V_{00}}$, for the 2007,
    2009.95, 2009.96, and 2009.97 epochs, as well as
    when all epochs are combined.  The minimum $\chi^2$, number of visibility
    observations ($N$) and fit parameters ($k$) are listed in each plot.
    In all cases the color map ranges from $\min(\chi^2)$ (blue) to
    $\min(\chi^2)+6$ (red).}\label{fig:agchi2}
\end{figure*}
During the 2007 epoch the mm-VLBI visibilities are concentrated nearly
exclusively along a single line, oriented nearly east-west (see the
top-left panel of Figure \ref{fig:Vobs}).  However, upon including the
2009 epochs, the portion of the $u$--$v$ plane sampled covers a
bow-tie shaped region with opening angle $26^\circ$ (see the
bottom-center panel of Figure \ref{fig:Vobs}).  While this is
insufficient to generate an image directly, the coverage is sufficient
to address the gross angular structure of Sgr A*'s image.  Where the
symmetric gaussian provides a phenomenological way in which to
estimate the typical size of Sgr A*'s emitting region, an asymmetric
gaussian can begin to probe its symmetry.

Figure \ref{fig:agchi2} shows the minimum $\chi^2$ (or equivalently,
the maximum likelihood) as a function of the average size, $\sigma$,
and anisotropy parameter, $A$.  The four left panels show this for the
individual epochs, while the large right 
panel shows this for the combined data set.  Unlike the symmetric
gaussian model, the likely regions have somewhat different
morphologies.  This is due to the different coverage of the $u$--$v$
plane during the various epochs (for example, the likely regions for
epochs 2007 and 2009.95 are similar because the $u$--$v$ coverage
during these epochs is similar).  Nevertheless, the region preferred
by the combined data sets is present in all cases, implying that as
with the symmetric gaussian all epochs are consistent with a single
underlying image morphology.  During this time the flux normalization
of the compact component varied from $2.23\,\Jy$ (2009.96) to
$3.06\,\Jy$ (2009.97).

\begin{figure*}
  \begin{center}
    \includegraphics[width=\textwidth]{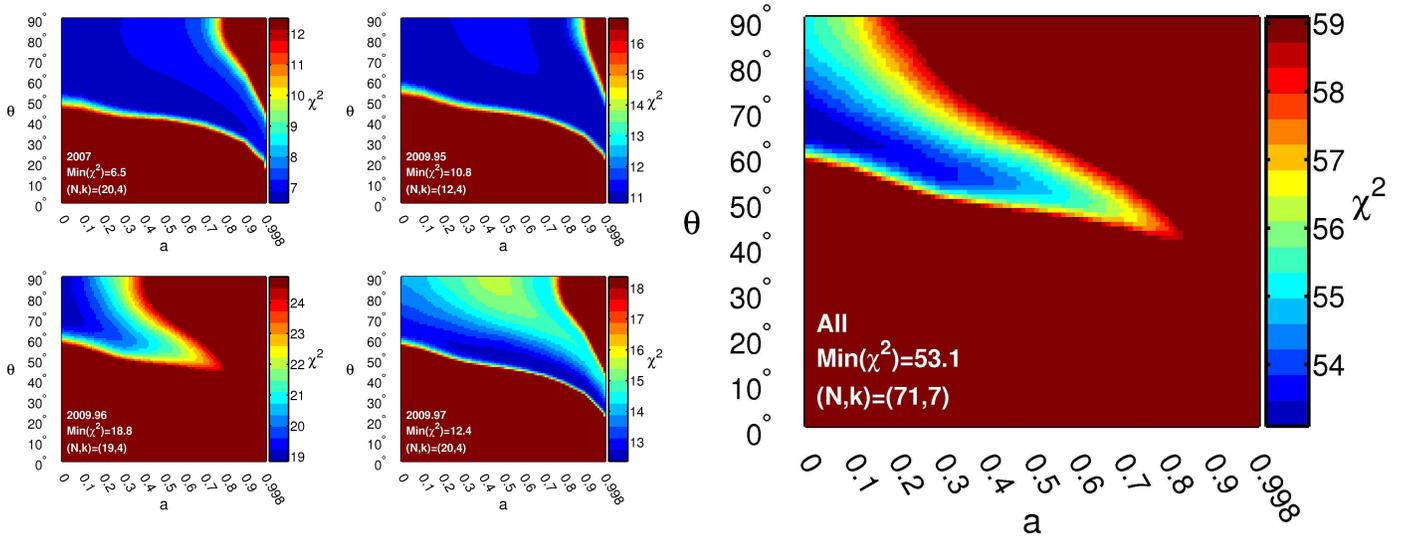}
  \end{center}
  \caption{BL06 accretion flow $\chi^2$'s as a function of $a$ and
    $\theta$, at the most likely $\xi$ and $\bmath{V_{00}}$, for the
    2007, 2009.95, 2009.96, and 2009.97 epochs, as well as
    when all epochs are combined.  The minimum $\chi^2$, number of visibility
    observations ($N$) and fit parameters ($k$) are listed in each plot.
    In all cases the color map ranges from $\min(\chi^2)$ (blue) to
    $\min(\chi^2)+6$ (red).}\label{fig:adchi2} 
\end{figure*}
For all epochs the reduced-$\chi^2$ is nearly unity, ranging from $0.45$
(2007) to $1.34$ (2009.95).  The most likely configuration is highly
asymmetric, with $\sigma=20.5^{+0.3+0.5}_{-0.8-1.3}\,\muas$,
$A=0.70^{+0.03+0.05}_{-0.1-0.18}$, and
$\xi={-19^\circ}^{+3^\circ+6^\circ}_{-1^\circ-2^\circ}$,
corresponding to a major--minor axis ratio of more than
$2.4^{+0.2+0.3}_{-0.4-0.6}$, with symmetric models highly disfavored.
The resulting ${\rm FWHM}$s of the minor and major axes are then
$37\pm1\,\muas$ and $88\pm9\,\muas$, though these are significantly
correlated due to the substantially larger fractional error on $A$ in
comparison to that on $\sigma$.
The intrinsic image, scatter-broadened image and associated
visibilities of the most likely configuration is shown in the middle
row of Figure \ref{fig:bestfit}.

The $\chi^2$ for the combined data set is $61.5$, and much lower than
that found for the symmetric case.  As described in Section
\ref{sec:BDA}, a decrease in $\chi^2$ is expected given the two
additional addition of two parameters.  However, the various $\IC$s,
given in Table \ref{tab:results}, provide a means for identifying
significantly lower $\chi^2$.  The best fit asymmetric model has a
$\BIC$ that is 6.8 lower than the best fit symmetric model, and an
$\AIC$ that is 10.4 lower than the best fit symmetric model.  These
provide ``strong'' evidence against symmetric models for the image
of Sgr A*, ruling these out at $2.6\sigma$ ($\BIC$) and
$3.2\sigma$ ($\AIC$) levels, in terms of the relative significance.
That is, despite the limited visibility coverage in the $u$--$v$
plane, the existing mm-VLBI observations can conclusively detect
asymmetric structure in Sgr A*.

\subsection{Accretion Flow and Implications of Physics}
\begin{figure*}
  \begin{center}
    \includegraphics[width=\textwidth]{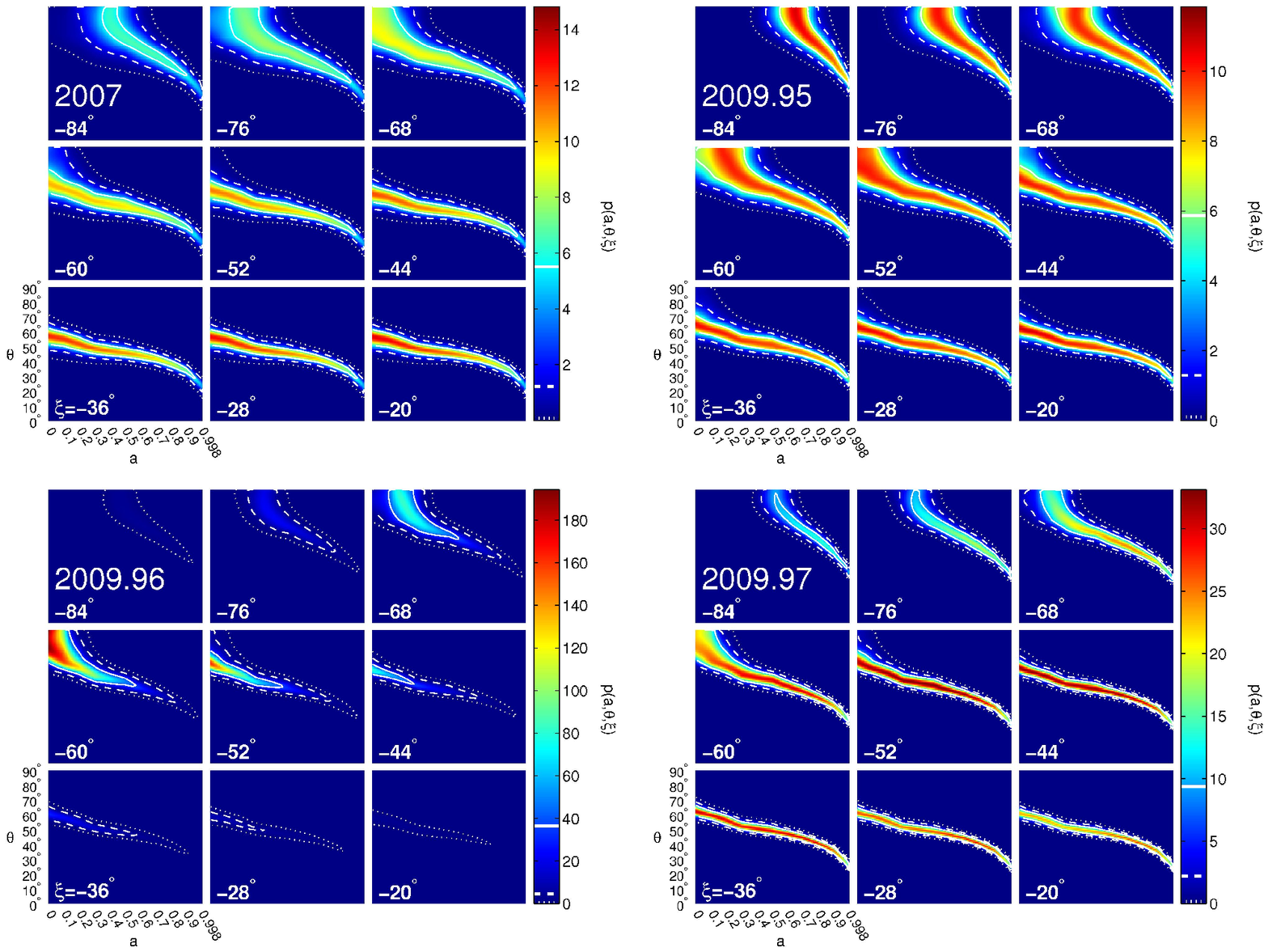}
  \end{center}
  \caption{Posterior probability of a given spin vector, marginalized
    over the $\bmath{V_{00}}$, obtained using the 2007, 2009.95,
    2009.96, and 2009.97 epochs.  Each sub-panel shows the posterior
    probability as a 
    function of $a$ and $\theta$ at fixed $\xi$, set to the value shown
    in the lower-left corner of each.
    In all cases the center panel shows the most probable
    position angle when all epochs are combined, and the color scheme
    spans identical ranges.  Contours demark the $1\sigma$
    (solid), $2\sigma$ (dashed), and $3\sigma$ (dotted) regions,
    defined by cumulative probability.}\label{fig:slice3Depochs}
\end{figure*}

The images of the BL06 model, described in Section \ref{sec:RIAF}, are
characterized by asymmetric crescents.  These arise due to the
combination of gravitational lensing, the relativistic orbital motion
and the opacity of the underlying accretion flow.  The size and extent
of the crescent depends upon the spin and inclination of the system,
with nearly face-on disks (small $\theta$) appearing annular.

Because we have only have access to the visibility magnitudes,
configurations rotated by $180^\circ$ are indistinguishable, imposing
an unavoidable ambiguity upon any results.  In addition, despite
opacity, the images exhibit nearly exact symmetry between configurations
viewed from above the equatorial plane (i.e., $\theta<90^\circ$) and
those viewed from below the equatorial plane (i.e.,
$\theta>90^\circ$) at equal inclinations.  As a consequence, there is
also an ambiguity in the line-of-sight component of the spin vector.
For this reason, here we discuss only $0^\circ\le\theta\le90^\circ$.
However we have performed the analysis for
$90^\circ\le\theta\le180^\circ$ as well, finding no statistically
significant differences in the parameter estimates.

Figure \ref{fig:adchi2} shows the minimum $\chi^2$ (maximum
likelihood) as a function of spin, $a$, and viewing angle, $\theta$.
As with Figure \ref{fig:agchi2}, the four panels on the left show this
for the individual epochs, while the large panel on the right shows
$\chi^2$ for the combined data set.  The morphology is similar in all
cases, with the different $u$--$v$ coverage during the different
epochs manifesting itself primarily in the size of the likely region.
As a result, we conclude that again the four epochs are consistent
with a single underlying image morphology.  Over all of the
observations the flux normalization varied from $2.16\,\Jy$ (2009.96)
to $3.00\,\Jy$ (2009.97), and is at all times sufficiently close to
the value of $2.5\,\Jy$ used to compute the images of the accretion
flow.  The intrinsic image, scatter-broadened image and corresponding
visibilities of the most probable configuration (not necessarily the
lowest $\chi^2$, see Section \ref{sec:EBHS}) is shown in the bottom
row of Figure \ref{fig:bestfit}.

At 53.09, the $\chi^2$ for the BL06 model (corresponding to a
reduced-$\chi^2$ of 0.830) is the smallest of any model we consider.
Again we may assess the significance of the this by appealing to the
$\IC$s described in Section \ref{sec:BDA}.  In this case the $\BIC$
and $\AIC$ are given by $82.9$ and $68.9$, respectively.  These are
much lower than those from the symmetric gaussian, providing
``decisive'' evidence against symmetric configurations.
Both are also below those of the asymmetric gaussian by $8.4$ (since
the number of parameters is the same in both models),
implying ``strong'' evidence for the physically motivated accretion
flow model in contrast to the phenomenological asymmetric gaussian.
This corresponds to a $2.9\sigma$ confidence level in terms of the
relative significance of the two models; i.e., the physically
motivated accretion models are more than 67 times as likely as the
phenomenological asymmetric gaussian models, and more than
$2\times10^3$ times as likely as symmetric gaussian models.  However,
for two reasons this actually understates the case.

First, the reduced-$\chi^2$ of the BL06 models is significantly
less than unity.  Indeed, with $64$ degrees of freedom we expect
a $\chi^2$ lower than that obtained (53.09) only $17\%$ of the time.
This suggests that the errors on the mm-VLBI visibilities have been
over-estimated by a moderate amount.  On the other hand, we may
measure the ``true'' errors by assuming that the BL06 model gives a
sufficiently close approximation to the true image flux by
renormalizing them until the reduced-$\chi^2$ is unity.  This requires
a roughly $10\%$ reduction in the errors quoted in Section
\ref{sec:SoO}.  This in turn alters the $\chi^2$, $\BIC$ and $\AIC$
for the other models as well.  These values are listed under the
``Implied Errors'' section of Table \ref{tab:results}.  The net effect
is to increase the significance with which the asymmetric gaussian is
ruled out in favor of the BL06 accretion model to $3.2\sigma$.

Second, the accretion flow models should generally be preferred on
the basis of their motivation alone.  That is, there is a strong prior
in favor of physically motivated models by virtue of their design and
connection to an existing body of knowledge.  Furthermore, in the case
of Sgr A* this includes the fact that these models were constrained to
fit the pre-existing spectral data as well, something that is only
possible because the physics governing the accretion flow provides a
means to relate the properties of images at different wavelengths.

The fact that the BL06 model provides a significantly better fit to
the mm-VLBI data implies that we can now distinguish phenomenological
and physically motivated models on the basis of mm-VLBI observations
alone.  Since the most prominent features of the accretion flow image
are due to the generic properties of black hole accretion flows,
namely, the spacetime and orbital motion, this is likely to be
robust among all radiatively inefficient accretion flow models for Sgr
A*.  That is, even with the extremely sparse $u$--$v$ coverage
presently available it is already possible to probe signatures of
general relativity and accretion physics in the image itself.

\section{Estimating Black Hole Spin} \label{sec:EBHS}
\begin{figure*}
  \begin{center}
    \includegraphics[width=\textwidth]{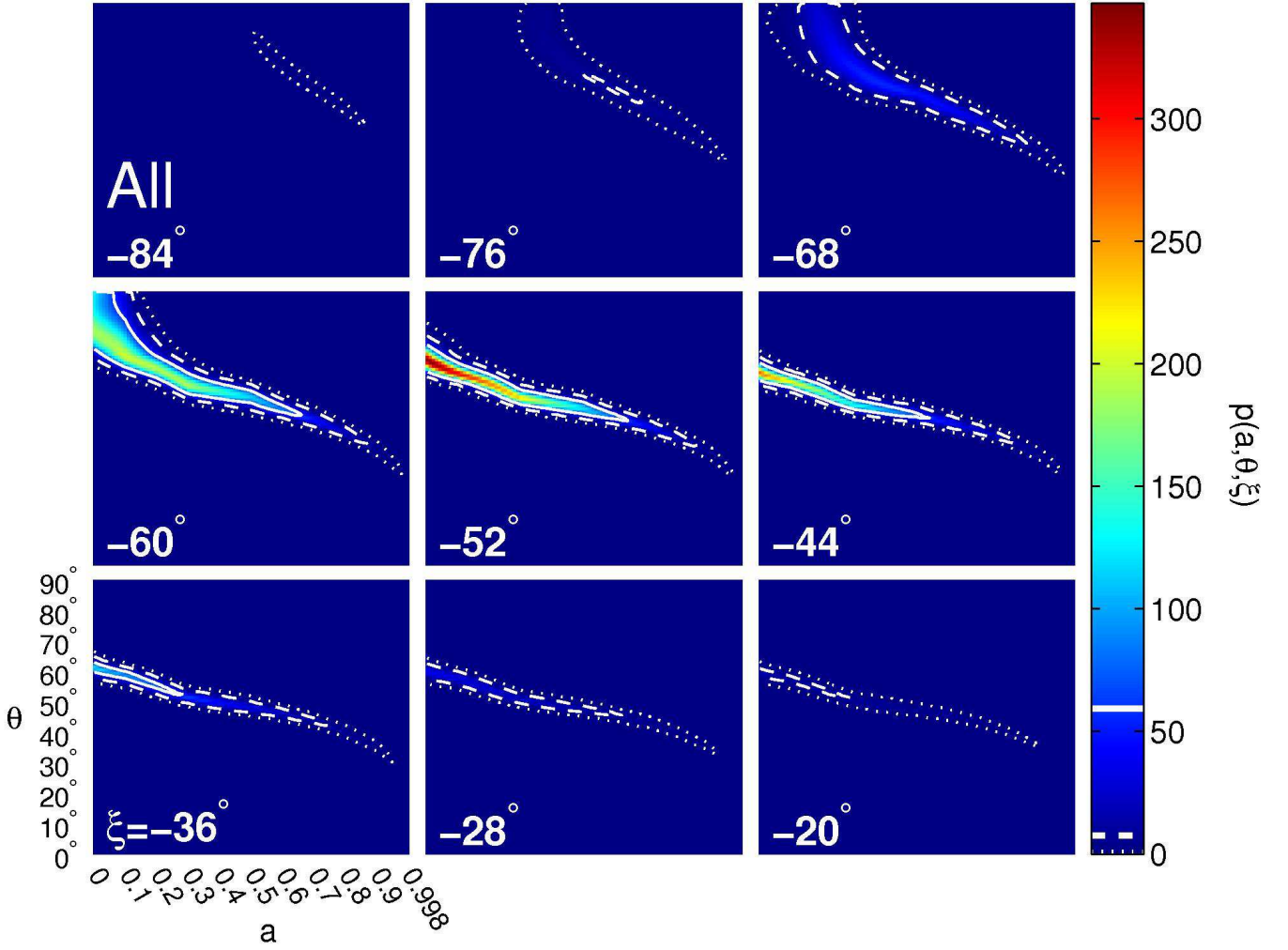}
  \end{center}
  \caption{Posterior probability of a given spin vector, marginalized
    over the $\bmath{V_{00}}$ when all epochs are
    combined.  Each sub-panel shows the posterior probability as a
    function of $a$ and $\theta$ at fixed $\xi$, set to the value shown
    in the lower-left corner of each.
    In all cases the center panel shows the most probable
    position angle when all epochs are combined, and the color scheme
    spans identical ranges.  Contours demark the $1\sigma$
    (solid), $2\sigma$ (dashed), and $3\sigma$ (dotted) regions,
    defined by cumulative probability.}\label{fig:slice3D}
\end{figure*}

Following \citet{Brod_etal:09} we produce posterior probability
distributions for the parameters defining the BL06 model, assuming a
flat prior on $a$ and isotropic priors upon the spin direction, and
marginalizing over the $\bmath{V_{00}}$.  For
the 3-dimensional parameter space defining the vector black hole spin,
$(a,\theta,\xi)$, this is done for each epoch individually (Figure
\ref{fig:slice3Depochs}) as well as for the combined data set (Figure
\ref{fig:slice3D}).  For each epoch we show the probability
distribution for the same position angle slices, defined such that the
most probable values from the combined data set is exhibited in the
central panel.  In all cases we define $1\sigma$, $2\sigma$ and
$3\sigma$ contours in terms of the cumulative probability, and
these are shown by the solid, dashed and dotted lines, respectively.
The probabilities are normalized to their average value, i.e., if all
points were equally probable the probability density would be unity.

Note that the way in which we have chosen the slices in $\xi$ in
Figure \ref{fig:slice3Depochs} does not capture the most likely
configuration based upon the 2007 epoch alone \citep[shown in
][]{Brod_etal:09}.  Nevertheless, the regions shown are well within
the $1\sigma$ region from that epoch.  Here we explicitly see that
all epochs produce consistent estimates for the spin, with varying
degrees of statistical strength.  

The combined data set dramatically restricts the parameter estimates
to a narrow sliver in the 3-dimensional spin parameter space.  The
most probable values for the spin are $a=0.0^{+0.64+0.86}$,
$\theta={68^\circ}^{+5^\circ+9^\circ}_{-20^\circ-28^\circ}$,
$\xi={-52^\circ}^{+17^\circ+33^\circ}_{-15^\circ-24^\circ}$, where the
errors quoted are the $1\sigma$ and $2\sigma$
errors.  At this point the probability density is roughly 350
times that of the average value.  In practice, these quantities are
much more tightly correlated with
\begin{equation}
\theta\simeq 68^\circ - 42^\circ a \pm 3^\circ \pm 5^\circ
\end{equation}
Note that these are degenerate with solutions for which $\xi$ differs
by $180^\circ$, i.e.,
$\xi={128^\circ}^{+17^\circ+33^\circ}_{-15^\circ-24^\circ}$, and for
which the line-of-sight component of the spin is reversed, i.e.,
$\theta={112^\circ}_{-5^\circ-9^\circ}^{+20^\circ+28^\circ}$.

We do not attempt to determine the systematic uncertainties associated
with selecting a particular accretion model.  However, we note that a
number of efforts to fit alternative accretion flow models to the 2007
epoch have reached consistent results despite differences in the models,
suggesting that these results are robust.  Furthermore, the quality of
the fits to the mm-VLBI visibility and spectral data, concurrently,
suggests that the features of the BL06 model responsible for
determining the spectral and image properties are generic, and are
therefore insensitive to the particulars of the accretion flow
models.  However, full studies of the systematic errors associated
with the particular choices made for the accretion flow properties and
the underlying spacetimes are now justified.  While beyond the purview
of the present paper, we will report upon such efforts elsewhere.

\begin{figure*}
  \begin{center}
    \includegraphics[width=\textwidth]{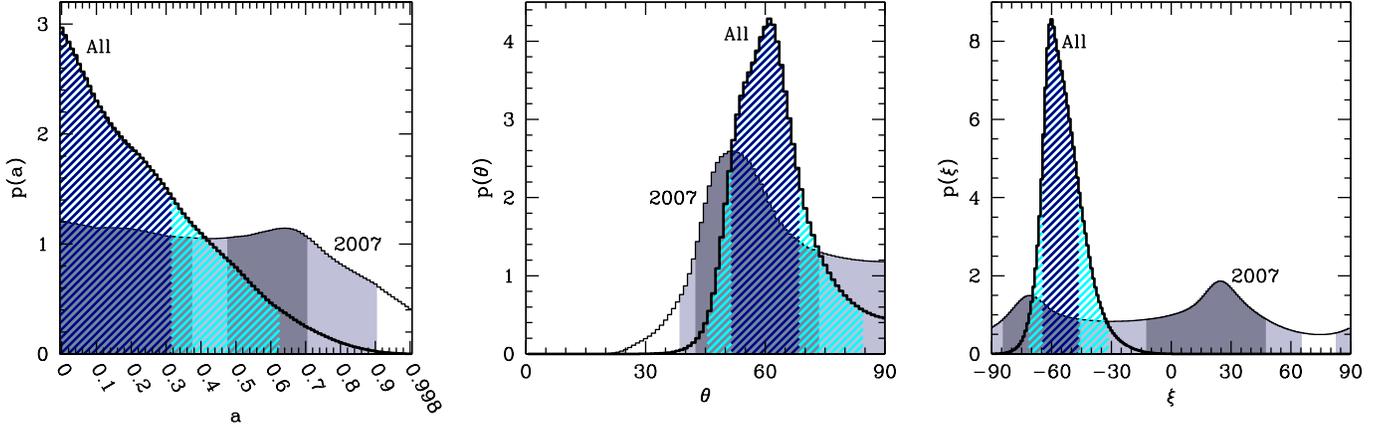}
  \end{center}
  \caption{Posterior probability of a given $a$ (left), $\theta$
    (middle), and $\xi$ (right), marginalized over all other
    parameters, obtained using all epochs.  The filled regions show the
    $1\sigma$ (dark) and $2\sigma$ (light) regions, defined by
    cumulative probability.}\label{fig:marg1D}
\end{figure*}

Probability distributions for each of the spin parameters,
marginalized over all others, are shown in Figure \ref{fig:marg1D}.
In addition to the present case, these are also shown for the analysis
of the 2007 epoch for comparison.  In these the $1\sigma$ and
$2\sigma$ ranges, defined by the cumulative probability, are also
shown.  In all cases the marginalized probability distributions from
the combined data set are much more narrowly peaked than their 2007
epoch counterparts.  Nevertheless, they are all consistent at the
$1\sigma$ level with those obtained from the 2007 epoch alone.

It is now possible to exclude $a>0.62$ at the $2\sigma$ level, with
$a=0^{+0.32+0.62}$, substantially preferring non-spinning models.  Thus
the high-spin island seen in Figure 7 of \citet{Brod_etal:09} is now
eliminated.  Similarly, the position angle 
is now very clearly constrained, choosing the solution less favored by
the 2007 epoch data (though still consistent at the $1\sigma$
level).  In this case we have $\xi={-60^\circ}^{+14^\circ+29^\circ}_{-5^\circ-12^\circ}$.
Finally, the most probable viewing angle is
$\theta={61^\circ}^{+7^\circ+24^\circ}_{-9^\circ-15^\circ}$.  This is
somewhat higher than the most likely value from the 2007 epoch alone,
though well within the $1\sigma$ uncertainty.

These estimates for the orientation of the spin vector are in good
agreement with a number of other efforts to estimate the properties of
Sgr A*'s accretion flow.  Estimates based upon fitting longer
wavelength observations with numerical models of radiatively
inefficient accretion flows produce position angles and inclination
estimates with large uncertainties, though these are nevertheless
consistent with the results obtained here
\citep{Huan-Cai-Shen-Yuan:07}.
It is also in excellent agreement with more recent attempts to probe
the spin orientation using the mm-VLBI data from the 2007 epoch
\citep{Huan-Taka-Shen:09,Dext-Agol-Frag-McKi:10}.
As before it is not possible to assess consistency with models that
employ qualitatively different plasma distributions near the black
hole \citep[e.g., ][]{Mark-Bowe-Falc:07}, though they tend to imply
similarly large viewing angles.

We find similar spin orientations to those inferred from modelling of
infrared polarization observations of Sgr A*'s flaring emission,
though in this case the uncertainties are considerable \citep[e.g.,
][]{Meye_etal:07}.  Similarly, we find consistency with the spin
directions obtained from modeling the spectrum and polarization using
general relativistic MHD simulations, despite preferring significantly
smaller spin magnitudes \citep{Shch-Penn-McKi:10}.

Unlike the estimates in \citet{Brod_etal:09}, there is no longer any
allowed solution for the spin vector that aligns with either of the
reported stellar disks in the inner $0.2\,\pc$ of the Galactic
center \citep{Genz_etal:03}.  However, our revised position angle
estimates are consistent with being aligned with the X-ray feature
reported in \citet{Muno_etal:08}, bolstering the interpretation of
this as related to a possible jet.  Note, however, this interpretation
may be inconsistent with the low spin magnitudes we prefer.

\section{Optimizing Future Observations} \label{sec:OFO}
\begin{figure*}
\begin{center}
\includegraphics[width=\textwidth]{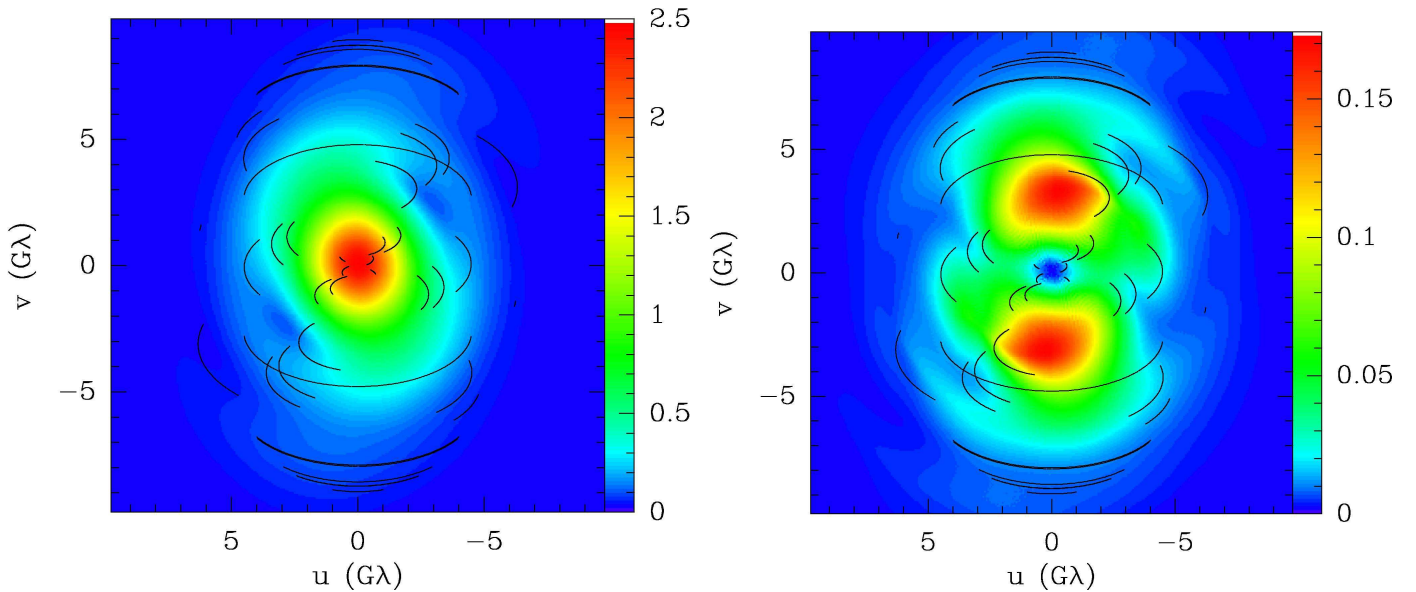}
\end{center}
\caption{Left: Mean predicted visibility amplitudes in Jy.  Right:
  Standard deviation of predicted visibility amplitudes.  In both
  cases the model image was weighted by its posterior
  probability.} \label{fig:heat}
\end{figure*}
The constraints upon the accretion model parameters obtained in the
previous sections have implications for future mm-VLBI experiments.
With these it is possible both to make predictions for the expected
visibilities on the various possible baselines, as well as identify
which baselines are most likely to provide substantial improvements to
the BL06 model parameter estimation.  To estimate these, here we
compute the average visibility amplitudes as well as the variance
associated with the uncertainty in the model parameters, weighted by
the posterior probability distributions we have obtained using the
combined mm-VLBI data set, following the method of \citet{Fish-Brod-Doel-Loeb:09}.

The probability-weighted mean visibility profile of the
scatter-broadened $230\,\GHz$ emission from Sgr A* is elongated in the
$(+u,+v)$ direction (Figure \ref{fig:heat}).  For the moderately high
values of $\theta$ favored by the mm-VLBI data, the intensity profile
is dominated by Doppler-boosted emission on the approaching side of the
accretion flow (e.g., in the northeast of lower left panel of Figure
\ref{fig:bestfit}).  This portion of the emission is elongated parallel to
the projected direction of the black hole spin vector.  Since the
effective size of the emission is larger along the projected spin
axis, the correlated flux density falls off faster with baseline
length for baselines that are sensitive to structure in this direction
than in the perpendicular direction.  Our estimates of RIAF parameters
suggest that a mm-VLBI baseline oriented southwest-northeast will
detect more correlated flux density than an equal-length baseline
oriented southeast-northwest.

The standard deviation of the visibility amplitudes provides an
estimate of which baselines would provide maximal additional
constraints on RIAF model parameters (assuming equal sensitivity at
all sites).  Previous computations based on the 2007 epoch of data
indicated that the largest scatter occurred at baseline lengths of
approximately $3\,{\rm G}\lambda$ and with orientations perpendicular to the
Hawaii-SMT baseline \citep{Fish-Brod-Doel-Loeb:09}.  These findings still hold in
light of the 2009 data.

Among possible observing baselines in the next few years, Chile-SPT
and Chile-LMT probe the region of highest standard deviation, followed
by baselines between the LMT and the continental US.  Our model
suggests that the LMT-SMT and LMT-CARMA baselines will detect well
over $1\,\Jy$.  The increased sensitivity provided by phased ALMA may be
important on the baselines to Chile, as the probability-weighted mean
visibility amplitudes are $\gtrsim 0.1\,Jy$ on the Chile--SMT baseline,
$\lesssim 0.1\,\Jy$ over most of the $(u,v)$ track of the Chile--CARMA
baseline, and smaller still on the longer baselines to Chile.
However, the standard deviation of the predicted model visibility
amplitudes is several $\times 10\,\mJy$ on these baselines, leading to
uncertainties of tens of percent in model amplitude predictions.
Further mm-VLBI data, either in the form of higher sensitivity on
existing baselines or detections on new baselines, will both reduce
these uncertainties and test the RIAF model with increasing rigor.

\section{Conclusions} \label{sec:C}

The significantly increased number of long-baseline visibilities,
significantly larger signal-to-noise, and improved north-south
coverage of the 2009 mm-VLBI observations have already paid
substantial dividends in the estimation of the properties of Sgr A*.
This is despite the fact that only three independent mm-VLBI stations 
(\JCMT,\CARMA,\SMT) were used, and therefore the $u$--$v$ plane
remains extremely sparsely populated, with long baselines primarily in
the east-west direction.  Constraints upon the black hole spin have
improved dramatically in all cases, with the spin and viewing angle
becoming tightly correlated, with the most probable configuration,
$a=0.0^{+0.64+0.86}$,
$\theta={68^\circ}^{+5^\circ+9^\circ}_{-20^\circ-28^\circ}$,
$\xi={-52^\circ}^{+17^\circ+33^\circ}_{-15^\circ-24^\circ}$
being roughly 350 times as likely as the average probability
density, and 25 times as likely as the most probable configuration
reported in \citet{Brod_etal:09}.

Despite the limited north-south coverage, the 2009 mm-VLBI data
conclusively excludes symmetric gaussian models for the source.  The
preferred asymmetric gaussian has a major--minor axis ratio of
$2.4^{+0.2+0.3}_{-0.4-0.6}$, oriented with the major axis oriented
${-19^\circ}^{+3^\circ+6^\circ}_{-1^\circ-2^\circ}$ east of north.
This implies major and minor axis ${\rm FWHM}$s of $37\pm1\,\muas$ and
$88\pm9\,\muas$, with the symmetric case excluded at $3.9\sigma$
significance.  Note that this orientation is not aligned with any
particular feature in Sgr A*'s vicinity or the properties of the
intervening interstellar electron scattering screen.

It is natural to give physically motivated models a prior bias over
phenomenological models.  Nevertheless, even when physically motivated
accretion models are compared with asymmetric gaussian models are
weighted equally, the accretion models provide a significantly better
fits to the mm-VLBI data.  Based upon both the $\BIC$ and $\AIC$ we
find strong evidence in favor of the accretion model, corresponding to
a posterior probability $67$ to $160$ times larger than that of the
most likely asymmetric gaussian model.  This is particularly striking
given the simplicity of the accretion model we consider, suggesting
that the image depends primarily upon the gross dynamical and
geometric properties of the system: the orbital motion of the
accreting material and the strong gravitational lensing by the black
hole.  In any case, it is clear that we have now entered the era of
studying accretion and black hole physics with mm-VLBI.

Our best fit accretion model requires a black hole spin of
$a=0.0^{+0.64+0.86}$ viewed at an angle of
$\theta={68^\circ}^{+5^\circ+9^\circ}_{-20^\circ-28^\circ}$, oriented
at a position angle of
$\xi={-52^\circ}^{+17^\circ+33^\circ}_{-15^\circ-24^\circ}$.
Ambiguities due to the fact that only amplitudes of the visibilities
were measured and the near symmetry of the accretion model images
produce degeneracies corresponding to
$\theta\leftrightarrow180^\circ-\theta$ and
$\xi\leftrightarrow\xi+180^\circ$, independently.  The detection of
closure phases will eliminate the ambiguity in position angle.

During all of the mm-VLBI epochs presently available, three or fewer VLBI
stations were employed, providing at most three baselines.  In
practice the two long baselines are nearly collinear, aligned
predominantly east-west.  The resulting sparse coverage within
$u$--$v$ plane is the primary factor limiting the estimation of black
hole and accretion flow parameters.  Therefore, despite the success
attained thus far, there is a considerable opportunity to improve the
constraints upon Sgr A* and its accretion flow dramatically in the
near future by including additional VLBI stations.
The {\em Event Horizon Telescope} is a $\mm$ and sub-$\mm$ wavelength VLBI
network whose goal is to observe, image, and time resolve structures
near the black hole event horizon \citep{Doel_etal:09b}.  Over 
the next few years, new sites will join the current array, and
sensitivities of critical baselines (e.g., those found in Section
\ref{sec:OFO}) will be enhanced through technical developments to widen
bandwidths and phase together multiple dishes at $\mm$ array sites (e.g.,
\CARMA, \SMA, \ALMA).  An important result is that within the next few
years, interferometric phase information on long VLBI baselines will
become available through robust measurement of the closure phase
\citep{Doel_etal:09}, which will provide important new constraints on
models of Sgr A*.

\acknowledgments
This work was supported in part by NSF grants AST-0907890, AST-0807843, and
AST-0905844, and NASA grants NNX08AL43G and NNA09DB30A.

\bibliography{gcpe2.bib}
\bibliographystyle{apj}

\end{document}